\documentclass[10pt]{iopart}               

\input{epsf}

\newcommand{\TKT}{T_{\scriptscriptstyle \rm KTB}}

\begin{document}
                 
\maketitle

\title[Vortex corrections to universal scaling of magnetic fluctuations 
in 2D XY model]{Vortex corrections to universal scaling of magnetic
fluctuations in 2D XY model}

\author{Peter C W Holdsworth$^1$\footnote{pcwh@ens-lyon.fr} and 
Mauro Sellitto$^2$\footnote{sellitto@ictp.trieste.it}}

\address{$^1$ Laboratoire de Physique, Ecole Normale Sup\'erieure de Lyon, 
\\ 46 All\'ee d'Italie, F-69364 Lyon cedex 07, France}

\address{$^2$ The Abdus Salam International Centre for Theoretical
  Physics, \\ Strada Costiera 11,  I-34100 Trieste, Italy}

\begin{abstract}
  The vortex contribution to the probability density function of
  longitudinal magnetization fluctuations is examined in finite
  2D XY systems close to the Kosterlitz-Thouless-Berezinskii
  transition temperature.  Within the temperature range studied their
  relevance is limited to rare fluctuations, where they increase the
  probability of events exceeding four standard deviations below the
  mean magnetization.
\end{abstract}


\bigskip

The characterization of fluctuation statistics is a central problem in
the study of critical phenomena, as the break down of Landau theory on
approaching
the critical point, implies a non-Gaussian distribution for
order parameter fluctuations~\cite{wil.74}.  From renormalization
group theory it is customary to think of critical phenomena divided
into universality classes, characterized by the symmetry group of the
order parameter and the spatial dimension.  One would therefore expect
critical fluctuation statistics to be determined essentially by the
set of critical exponents (say $\beta$ and $\nu$ for a regular
critical point) which describe the scaling behavior of derivatives of
the singular part of the free energy. Evidence from, e.g. the Ising
model~\cite{Bruce,binder} and Potts models~\cite{Botet} suggests that
this is indeed generally the case.

We have, however recently considered an exception to this established
phenomenology, the 2D XY model~\cite{bra.01,por.01}.  Here the low
temperature phase consists of a line of critical points, with one
independent critical exponent $\eta(T) = T/2\pi J$ (with $J$ the
exchange constant) extending down to zero temperature and separated
from the high temperature paramagnetic phase by the
Kosterlitz-Thouless-Berezinskii (KTB) phase
transition~\cite{kos.73,kos.74,ber.71}. It is well established that
for all temperatures below the KTB transition temperature, $\TKT$,
renormalisation group flows are to a quadratic effective
Hamiltonian~\cite{vil.75,jos.77}, with the result that the asymptotic
behaviour, in the limit of large system size is perfectly captured by
a harmonic model. The advantage of such a simple model is that the
probability density function, $P(m)$, for fluctuations of the order
parameter, $m$, can be calculated analytically, without using either
renormalisation group or the scaling hypothesis.  Surprisingly, we
find that the form of distribution is independent of $\eta$ along the
whole line of critical points. This universal scaling function,
plotted as the solid line in Fig.~1, arises when $\sigma P(m)$ is
plotted against $\mu=(m-\langle m \rangle)/\sigma$, where $\langle m
\rangle$ and $\sigma$ are the mean and standard deviation of the
distribution~\cite{bra.98,bra.01,aji.01}. It is asymmetric, with an
exponential tail for fluctuations below the mean and double
exponential for fluctuations above the mean.

It has been observed that very similar distributions occur in two and
three dimensional Ising models at a temperature $T^{\ast}(L)$
slightly below, but close to the critical
temperature~\cite{bra.00,comment}. We have argued that this is a
critical phenomenon~\cite{bra.00,reply}, giving weight to our earlier
proposition~\cite{bra.98} that many correlated systems, both in and
out of equilibrium, can be driven into a state with very similar
fluctuations~\cite{bra.00,LPF}, irrespectively of their universality class.

In this Letter we examine the vortex corrections to this universal
scaling function as one approaches $\TKT \simeq
0.8929(1)$~\cite{TKT1,TKT2}.  We confirm, on the one hand, that below
$\TKT$ the effect of vortex pairs will disappears as the thermodynamic
limit is taken, while above $\TKT$ they become dominant in the same
limit. On the other hand, we show that their relevance changes only
very slowly with length scale, with the result that finite size
corrections to the thermodynamic limit distribution function will be
observable over a large range of system sizes.

\smallskip

The 2D XY model is defined by the Hamiltonian
\begin{eqnarray}
\label{H}
H =-J\sum_{\langle i,j \rangle }\cos(\theta_i-\theta_j) \,,
\end{eqnarray}
where the angle $\theta_i$ gives the orientation of a classical spin
vector of unit length, confined to a plane (the sum being over nearest
neighbors spins). In the following we consider a square lattice of
side $L$ with periodic boundaries, and we set throughout
$J/k_{\scriptscriptstyle \rm B}=1$. We define the magnetization $m$
for a single configuration as
\begin{eqnarray}\label{eq-mag}
m = \frac{1}{N} \sum_{i=1,N} \cos(\theta_i - \overline{\theta}) \,,
\end{eqnarray}
where $\overline{\theta} = \tan^{-1}\left(\sum_i \sin \theta_i /\sum_i
\cos \theta_i\right)$ is the instantaneous magnetization direction.
As the physics of the low temperature phase is perfectly captured by a
harmonic, or spin wave Hamiltonian~\cite{vil.75,jos.77}, one can,
without loss of generality, develop the cosine interaction to 
second order and neglect the periodicity of $\theta_i$. This
Hamiltonian is diagonal in reciprocal space and can be solved
straightforwardly. In Refs.~\cite{bra.01,por.01} we find analytically
that $\sigma P(m)$, shown as the solid line in Fig. 1, is a universal
scaling function, not only of system size, but also of temperature and
therefore of critical exponent $\eta$.
\begin{figure}
  \epsfxsize=3.4in \centerline{\epsffile{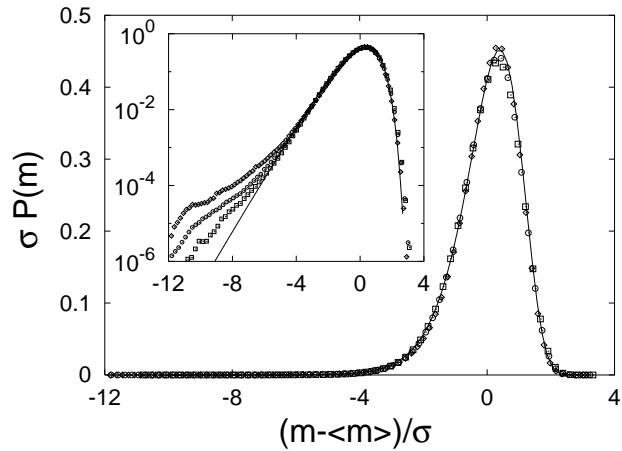}}
  \caption{Probability distribution function of magnetization $m$ in a
  2D XY system of size $L=32$, at temperature $T$ around the $\TKT$
  transition: $T=0.86 \,(\sqcup \!\! \!\! \sqcap),\, 0.89\,
  (\circ),\,0.92\,(\diamond)$. In the inset the plot is shown in
  semi-log scale. The full line represents the exact solution of the
  harmonic 2D XY model.}
\label{P_L32_T}
\end{figure}  

We stress that this analytic result is entirely due to harmonic spin
waves.  At any temperature, corrections to it come from two sources:
anharmonic, but analytic terms in the expansion of Eq. (\ref{H}) and
vortex pairs excited as the full periodicity of $\theta_i$ becomes
important in the region of $\TKT$~\cite{arc.97}.  The former have been
discussed in detail in Ref.~\cite{bra.01,lab.00}; at fixed
temperature, they give small deviations from the universal curve for
small system size, $L$.  These are finite size corrections to the
thermodynamic limit function and disappear with relatively modest
increase in the system size.

The two effects can be separated as vortices appear in appreciable
numbers in a small range of temperature close to $\TKT$
only~\cite{jen.91}. In order to study the effect of vortices on the
probability density function (PDF) we have carried out extensive Monte
Carlo (MC) simulations of the 2D XY model in a small range of
temperature around $\TKT$ and for different system sizes.  We
typically use systems of linear size $L=16,\,32,\,64$.  The system was
first equilibrated for $10^5-10^6$ MC sweeps (MCs), the probability
distribution of magnetization was then computed along a trajectory of
$10^7-10^8$ MCs, according to the size of the system.

In Fig.~\ref{P_L32_T} we show results for a fixed system size ($L=32$)
at several temperatures around $\TKT$.  First we observe that, when
numerical data are plotted in natural units, there is a rather good
qualitative agreement with the theoretical curve; a more careful
inspection however reveals systematic deviations in the tails: when
observed on a semi-logarithmic scale the exponential tail of the
distribution changes dramatically, large deviations below the mean
magnetization becoming more probable below a characteristic breaking
point. This is the vortex contribution. The break point shifts to
higher probability with increasing temperature, consistently with the
fact that the vortex density increases around $\TKT$. However at fixed
system size there is no dramatic difference in behaviour on crossing
$\TKT$.  The breaking point is a clear signature of two distinct
contributions to the PDF~\cite{arc.97}: small fluctuations are
dominated by spin-waves, while large ones by vortices. It is
interesting to note, however that, even at $\TKT$, the vortex
dominated region is limited to small probabilities and the range
$\left| m - \langle m \rangle \right| > 4\sigma$, making it difficult
to observe. For example Zheng and Trimper in Ref.~\cite{comment}
present numerical results concerning an XY system of size $L=32$ at
$T=0.89$; but in their plot deviations to the spin wave result do not
appear as their data is limited to the range $\left| m - \langle
m \rangle \right| \le 4 \sigma$. Deviations from the spin wave scaling 
function are also observed in Ref.~\cite{aji.01}, however the range of 
temperatures used is outside that for which one expects to see signature
of vortices for the probabilities resolved. Note that the best agreement
with the data originally published in Ref.~\cite{arc.97} is for the
highest temperature studied.

\begin{figure}
  \epsfxsize=3.4in \centerline{\epsffile{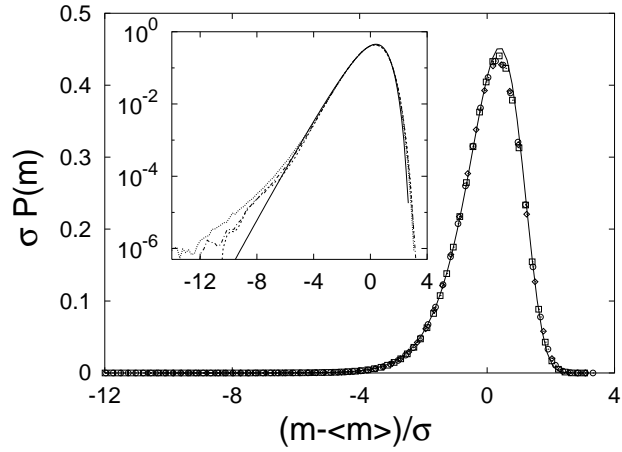}} \caption{PDF
  of magnetization $m$ in a 2D XY system of size $L=16 \,(\sqcup \!\!
  \!\! \sqcap),\, 32\, (\circ),\,64\,(\diamond)$, at temperature
  $T=0.86$ slightly {\sl below} the $\TKT$ transition.  In the inset
  the plot is shown in semi-log scale (dotted, dash-dotted, and dashed
  lines correspond to $L=16,\, 32,\,64$ respectively).}
\label{P_L_T086}
\end{figure}

Let us now consider how does the vortex contribution change as a
function of the system size. In Figs. 2 and 3 we present the PDF at a
fixed temperatures, slightly below and above $\TKT$.  In
Fig.~\ref{P_L_T086}, with $T=0.86$ one can see that, although the
effect of changing size is rather small, the deviation from the spin
wave result is reduced on increasing $L$, with the vortex dominated
region having a system size dependent slope.  This is again consistent
with a finite size correction to the thermodynamic limit function,
although it is clear that much bigger sizes would be required to
eliminate the effects of vortices over the range of probabilities
shown here.

\begin{figure}
  \epsfxsize=3.4in \centerline{\epsffile{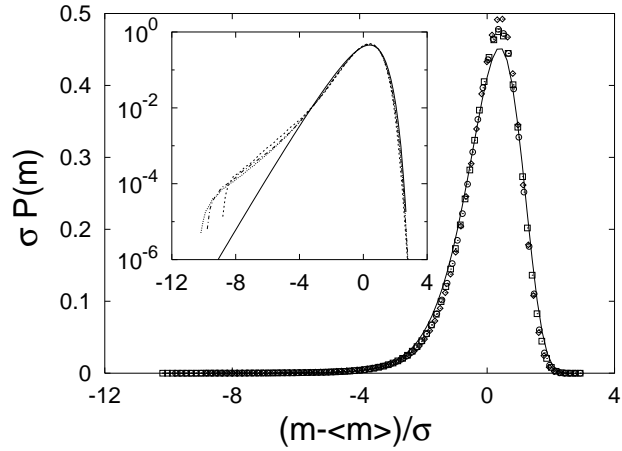}} \caption{PDF
  of magnetization $m$ in a 2D XY system of size $L=16 \,(\sqcup \!\!
  \!\! \sqcap),\, 32\, (\circ),\,64\,(\diamond)$, at temperature
  $T=0.95$ slightly {\sl above} the $\TKT$ transition. In the inset
  the plot is shown in semi-log scale (dotted, dash-dotted, and dashed
  lines correspond to $L=16,\, 32,\,64$ respectively).}
\label{P_L_T095}
\end{figure}

>From Fig.~\ref{P_L_T095} where $T=0.95$, one can see that above
$\TKT$ just the opposite happens; the slope in the vortex dominated
region increases slowly with increasing $L$. The region extends out to
a cut off, which decreases with system size.  The cut off appears
because, with fluctuations of increasing amplitude, the constraint $m
\ge 0$ comes into play, limiting the range of possible values of
$\mu$. For larger system sizes, or higher temperature the cut off
influences the form of the PDF, the topology changes and the system
enters the paramagnetic phase through vortex unbinding~\cite{arc.97}.

Since the spin-spin correlation length, $\xi$ diverges exponentially
with the approach to $\TKT$ from above~\cite{kos.74}:
\begin{equation}
\xi \approx \exp\left({\pi\over{\sqrt{c(T-\TKT)}}}\right),
\end{equation}
($c \approx 2$), finite size corrections to $\TKT$ are logarithmic in
$L$~\cite{bra.93,bra.94}. Defining $T_C(L)$ as the temperature where
$\xi = L$ gives
\begin{equation}
T_C(L) = \TKT + {\pi^2\over{c(\log L)^2}}.
\end{equation}
At least within finite size scaling terms, this gives an exceptionally
large shift. Our results are consistent with this, as at the
temperature shown in Fig.~\ref{P_L_T095} the correlation length is in
the range of system sizes studied~\cite{bra.93}.  The results beg the
question: can one observe scale independence for both the spin wave
and the vortex contribution to the PDF, giving a universal two
component scaling distribution?  Data collapse onto a single curve
along a locus of temperatures with varying system size, would be
consistent with the notion of the temperature $T^{\ast}(L)$,
introduced in Ref.~\cite{bra.93} and put on a more rigorous footing in
Ref.~\cite{Chung}. At $T^{\ast}(L)$, the effective coupling constant,
renormalized by the vortex pairs in the system, is independent of
length scale, over all lengths up to size $L$~\cite{bra.94}. It lies
in the interval $\TKT < T^{\ast}(L)< T_C(L)$ and also scales
logarithmically towards $\TKT$ with system size. It corresponds to a
situation where the vortex-vortex correlations are independent of
scale and one would therefore expect two component universality for
the PDF at this point.  
The logarithmic finite size scaling means that this is a
difficult question to address quantitatively and we have not attempted
to do so here.  However, we do show in Fig. 4 data collected at
$T=0.93$, for three system sizes, much smaller than the thermodynamic
correlation length.  We get fairly good data collapse and dependence
on $L$ is extremely small over the range studied. There is some
deviation from the common curve for $L=64$, and one could clearly
obtain better collapse by moving to a smaller temperature. This
indicates that at least one locus of points could exist giving data
collapse onto a single curve.

\begin{figure}
  \epsfxsize=3.4in \centerline{\epsffile{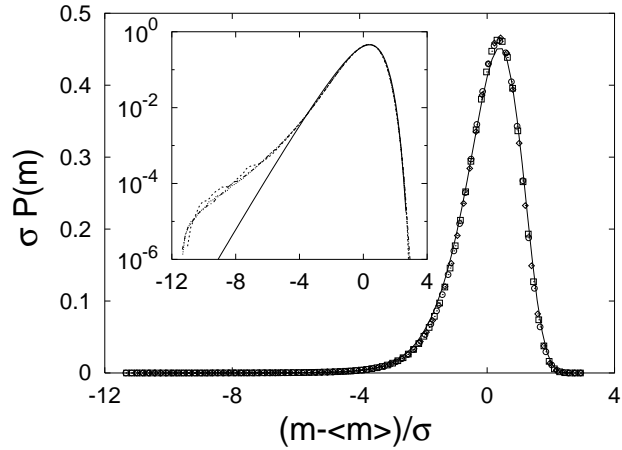}} \caption{PDF
  of magnetization $m$ in a 2D XY system of size $L=16 \,(\sqcup \!\!
  \!\! \sqcap),\, 32\, (\circ),\,64\,(\diamond)$, at temperature
  $T=0.93$. Inset: dotted, dash-dotted, and dashed lines correspond to
  $L=16,\, 32,\,64$ respectively.}
\label{P_L_T093}
\end{figure}  

In summary, we have presented a quantitative estimation of the vortex
contribution to the probability density function of longitudinal
fluctuations of magnetization in the 2D XY model, in a band of
temperature above and below the Kosterlitz-Thouless-Berezinskii
transition temperature.  The vortices influence the tail of the
distribution for large fluctuations below the mean, giving a break in
the exponential tail towards larger probabilities. In the range of
parameter space studied, the break is typically for fluctuations of
four standard deviations below the mean. Given the expected
logarithmic finite size scaling we suggest that this will always be
the case, within the physical domain probed numerically or
experimentally.

\bigskip

\ack

It is a pleasure to thank S.T. Bramwell and B. Portelli for
stimulating discussions.  MS acknowledges the support of CNRS
(contract N. 186078) and the
Laboratoire de Physique of ENS-Lyon, where this work was completed.

\bigskip

\section*{References}

\end{document}